\documentclass[conference]{IEEEtran}
\IEEEoverridecommandlockouts

\usepackage{subcaption}
\usepackage{bm}
\usepackage{siunitx}
\usepackage{algorithm}
\usepackage{algorithmic}

\usepackage{cite}

\usepackage{amsmath,amssymb,amsthm,textcomp}
\usepackage{enumerate}
\usepackage{multicol}
\usepackage{tikz}

\usetikzlibrary{positioning,chains,fit,shapes,calc}

\usepackage{algorithm}
\usepackage{algorithmic}

\usepackage{color}
\definecolor{myc1}{rgb}{0,0,0}

\begin{document}

\title{{HARQ Delay Minimization of 5G Wireless Network with Imperfect Feedback}  }

\author{
\IEEEauthorblockN{Weihang Ding\IEEEauthorrefmark{1} and Mohammad Shikh-Bahaei\IEEEauthorrefmark{1}
                  }
\IEEEauthorblockA{\IEEEauthorrefmark{1}Centre for Telecommunications Research, Department of Engineering, King's College London, London WC2R 2LS, UK.}

\vspace{-2em}
}

\maketitle

\begin{abstract}
5G new radio (NR) technology is introduced to satisfy more demanding services.
Ultra Reliable Low Latency Communication (URLLC)
requires very low delay compared with the previous techniques. This is hard to achieve when hybrid automatic repeat request (HARQ) is applied and especially when the feedback channel is erroneous. In this work, we consider various delay components in incremental redundancy (IR) HARQ systems and minimize the average delay by applying asymmetric feedback detection (AFD) and find the optimal transmission length for each transmission attempt.
A M/G/1 queuing model is used in this work to analyze the queuing delay in 5G NR when there are multiple uses in the system. Numerical results show that significant performance gains and lower outage probability can be achieved by applying AFD.

\end{abstract}

\IEEEpeerreviewmaketitle

\section{Introduction}

The fifth generation (5G) wireless mobile networks are developed to achieve substantial performance improvements \cite{Agiwal}. Low latency and high reliability are key requirements in most scenarios, particularly in Ultra-Reliable Low Latency Communication (URLLC). Ideally, 1ms end-to-end latency and 99.999\% reliability are expected, but there are still some hurdles to work out. For example, hybrid automatic repeat request (HARQ) acts as a bottleneck to the optimization of user-plane latency.

Combining Forward error correction (FEC) and automatic repeat request (ARQ), HARQ is a high-efficiency technique for data transmission, which performs much better than ordinary ARQ in poor channel conditions. Based on the soft-combining method at the receiver, HARQ can be divided into chase-combining HARQ (CC-HARQ) and incremental redundancy HARQ (IR-HARQ). In CC-HARQ, the same codeword is transmitted in different transmissions, while in IR-HARQ, retransmissions include additional parity-check bits. We only consider IR-HARQ in this work as coding gain is more significant in general wireless communication systems.
In the conventional IR-HARQ protocols, the decoding state is released by the receiver using feedback. If the feedback is binary, the transmitter is only able to know whether the decoding was successful, hence the length and power of each transmission attempt is fixed \cite{Jabi,Trillingsgaardems}. With longer feedback, the transmitter can acquire more information about the decoding states, and based on which, the transmitter might be able to adaptively select some of the parameters of the subsequent transmission. Adaptive system is a promising technique to enhance the performance of various systems by adjusting the system configuration based on the system parameters \cite{2,3,4,6,10,12}. Adaptive ARQ in cognative radio networks can increase the utilization of the resources, and somehow reduce overall delay \cite{12,13,14,15}. Similarly, adaptive HARQ is widely studied based on different transmission parameters, including rate adaptation \cite{Szczecinski}, \cite{Khosrvirad2014}, power adaptation \cite{Tuninetti}, \cite{Chaitanya}, and adaptive modulation and coding (AMC) schemes\cite{Choi}.

The feedback has to be prolonged to convey the full decoding state. In \cite{Jabi}, the relationship between feedback resolution and the overall throughput is studied. The results show that if the feedback includes more than 8 bits, the performance is very close to an ideal feedback HARQ system. However, transmitting 8-bit feedback is still unrealistic in reality due to the high cost of the feedback channel. When the quality of the feedback channel is low, applying extra feedback bits will increase the feedback error rate and make the error harder to analyse.

In most of the previous works, the feedback channel is assumed to be deterministic whereas the error probability of the HARQ feedback can not be made arbitrarily low in reality. A feedback error rate of 1\% is reasonable in LTE \cite{LTEchap12} and 0.1\%-1\% in 5G NR\cite{5GNR}. With limited resource \cite{1,11}, unreliable feedback will impair the performance of the HARQ system. There are only a few contributions on HARQ with unreliable feedback. In \cite{Breddermann2014}, \cite{Ahmad}, the feedback channel is modeled as a Binary Symmetric Channel (BSC). In \cite{Malak}, a notion of guaranteeable delay is introduced while the feedback channel is an erasure channel. Different from conventional symmetric feedback detection, asymmetric feedback detection (AFD) is introduced in \cite{Shariatmadari} to provide a better protection to negative acknowledgement (NACK) without assigning extra resources. In our previous work \cite{AFD}, we apply AFD to rate-adaptive HARQ and find the optimal thresholds and transmission rates to maximize throughput.

In this work, we study the long-term average delay of a HARQ process with imperfect feedback. We consider multiple delay components including queuing delay, transmission delay, and processing delay. We apply both AFD and rate-adaptive HARQ in this work to minimize the long-term average delay of the system. Another advantage of AFD is that it reduces the outage probability, so we also study the performance of the systems with outage probability constraints. Simulation results show that our proposed scheme can significantly reduce the long-term average delay compared with traditional methods.

\section{System model}

The round-trip delay (RTD) of the system is the time it takes for the packet to be sent and acknowledged. Among these delay components, the propagation delay and the transmission delay of the feedback are negligible compared with the other components. Therefore, the overall RTD can be expressed as the summation of the queuing delay, the transmission delay of the packet, and the processing delay at the receiver:
\begin{equation}
    T_{tot} = T_{queue}+T_{tran}+T_{proc}.
\end{equation}

Once a HARQ round begins, the source encodes the original packet of $N_b$ bits using a designated code rate. The feedback is generated by the destination according to the decodability of the packet. If the transmitter observes a NACK, it will transmit a certain amount of redundancy in the subsequent transmission attempt. The current HARQ round can only terminate as long as an ACK is observed, or the number of transmissions reached the maximum limit $M$.

\subsection{Channel model}
It is assumed that the packets are transmitted through block-fading channels with additive Gaussian noise. The $k$-th received symbol can be written as:
\begin{equation}
    \bm{y}_k=\sqrt{\text{SNR}_d}h_k\bm{x}_k+\bm{z}_k,
\end{equation}
where $\bm{x_k}$, $\bm{y_k}$, and $\bm{z_k}$ are the $k$-th transmitted symbol, received symbol and additive noise respectively, $h_k$ is the instantaneous channel fading coefficient, SNR$_d$ is the average signal to noise ratio of the downlink channel. 

We assume that the time between two transmission attempts is significantly larger than the coherence time, so 
$h_k$ is independent identically distributed and remains constant during a single transmission attempt. The transmitter has no knowledge about $h_k$ and is not able to predict it using the previous information before the transmission. $|h_k|$ is Rayleigh distributed with unity power gain $\mathbb{E}[|h_k|^2]=1$ and zero mean. $\bm{z}_k$ is a complex vector of zero mean, and unitary-variance Gaussian variables representing the noise. It is assumed that SNR$_d$ remains constant and is known by the transmitter. The decoding is based on the received sequence $\bm{y}$, and uses maximum likelihood decoding.

\subsection{Queuing model}

It is assumed that the packets arriving at the gNB follow a Poisson distribution. The system can be modeled as a M/G/1 queue, since packets are transmitted one-by-one, and the transmission time follows a general distribution based on the length of different transmission attempts.

Unlike LTE, 5G NR supports multiple subcarrier spacings from 15kHz up to 240kHz. Different subcarrier spacings correspond to different slot durations, as shown in Table \ref{subcarrier}.

\begin{table}[h!]
\centering
\caption{5G NR subcarrier numerology}
\label{subcarrier}
\begin{tabular}{|c|ccccc|}
\hline
$\mu$       & \multicolumn{1}{c|}{0}     & \multicolumn{1}{c|}{1}     & \multicolumn{1}{c|}{2}     & \multicolumn{1}{c|}{3}    & 4    \\ \hline
Sub-carrier spacing (\si{\kilo\hertz}) & \multicolumn{1}{c|}{15}    & \multicolumn{1}{c|}{30}    & \multicolumn{1}{c|}{60}    & \multicolumn{1}{c|}{120}  & 240  \\ \hline
Slot duration (\si{\micro\second})          & \multicolumn{1}{c|}{1000}  & \multicolumn{1}{c|}{500}   & \multicolumn{1}{c|}{250}   & \multicolumn{1}{c|}{125}  & 62.5 \\ \hline
Slots per subframe        & \multicolumn{1}{c|}{1}     & \multicolumn{1}{c|}{2}     & \multicolumn{1}{c|}{4}     & \multicolumn{1}{c|}{8}    & 16   \\ \hline
Symbols per slot          & \multicolumn{5}{c|}{14}                                                                                                 \\ \hline
OFDM symbol duration (\si{\micro\second})     & \multicolumn{1}{c|}{66.67} & \multicolumn{1}{c|}{33.33} & \multicolumn{1}{c|}{16.67} & \multicolumn{1}{c|}{8.33} & 4.17 \\ \hline
Supported for data        & \multicolumn{1}{c|}{Yes}   & \multicolumn{1}{c|}{Yes}   & \multicolumn{1}{c|}{Yes}   & \multicolumn{1}{c|}{Yes}  & No   \\ \hline
\end{tabular}
\end{table}

The transmission time (service time) of the $i$-th transmission attempt in a HARQ round $T_{i}$ can be calculated as $T_{i}=\frac{n_i}{14}T_{slot}$, where $n_i$ is an integer denoting the number of slots required to transmit this packet.

\section{Asymmetric feedback detection}

If the feedback channel is perfect, an outage occurs only when the receiver is still unable to recover the message after $M$ transmission attempts. If an outage occurs, this transmission will be handed over to the higher layer protocols and a new HARQ round will be initialized. Outages can cause significant system performance degradation. For normal HARQ processes, an outage probability of 0.1\%-1\% is tolerable \cite{5GNR}. If the feedback channel is imperfect, NACK errors will cause outages, and ACK errors will only lead to unnecessary retransmissions.

\begin{figure}[!t]
\centering
     \begin{subfigure}[b]{0.2\textwidth}
            \centering
            \resizebox{0.8\textwidth}{!}{
            \begin{tikzpicture}
            \fill [gray!1](-3,3) -- (-3,3) -- (-3,-3) -- (3,-3) -- (3,-3);
            \fill [gray!50](3,3) -- (-3,3) -- (3,-3);
            \draw [thick,->](-3,0) -- (3,0);
            \draw [thick,->](0,-3) -- (0,3);
            \draw [thin,black](-3,3) -- (3,-3);
            \draw [thick] (0,0) circle (2);
            \filldraw[black] (1.41421,1.41421) circle (2pt) node[anchor=west] {ACK};
            \filldraw[black] (-1.41421,-1.41421) circle (2pt) node[anchor=east] {NACK};
            \filldraw[black] (2.4,0.3) circle (0.1pt) node[anchor=east] {1};
            \end{tikzpicture}
            }
            \caption{SFD}
            \label{PSKsym}
     \end{subfigure}
     \begin{subfigure}[b]{0.2\textwidth}
            \centering
            \resizebox{0.8\textwidth}{!}{
            \begin{tikzpicture}
            \fill [gray!1](-2.3,3) -- (-3,3) -- (-3,-3) -- (3,-3) -- (3,-2.3);
            \fill [gray!50](3,3) -- (-2.3,3) -- (3,-2.3);
            \draw [thick,->](-3,0) -- (3,0);
            \draw [thick,->](0,-3) -- (0,3);
            \draw [thin,black](-2.3,3) -- (3,-2.3);
            \draw [thick] (0,0) circle (2);
            \draw [thin,<->] (0,0) -- (0.35,0.35) node[anchor=east] {$\alpha$};
            \filldraw[black] (1.41421,1.41421) circle (2pt) node[anchor=west] {ACK};
            \filldraw[black] (-1.41421,-1.41421) circle (2pt) node[anchor=east] {NACK};
            \filldraw[black] (2.4,0.3) circle (0.1pt) node[anchor=east] {1};
            \end{tikzpicture}
            }
            \caption{AFD}
            \label{Asym sch1}
     \end{subfigure}
     \caption{Constellation comparison between SFD and AFD.}
     \label{Const}
\end{figure}
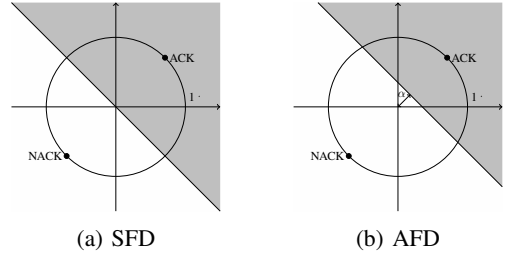

\begin{figure}[!t]
	\centering
	\includegraphics[width=0.85\columnwidth]{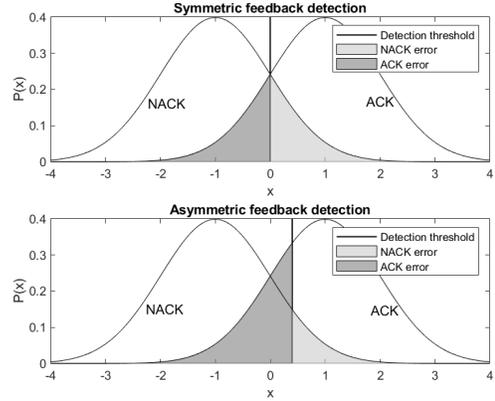}
	\caption{Comparison between SFD and AFD}
	\label{Compare_pdf}
\end{figure}

The feedback is binary quantized based on a feedback detection threshold. In symmetric feedback detection (SFD), the feedback detection threshold is set as Fig. \ref{PSKsym} so that the areas of ACK and NACK on the constellation are equal. However, in the case of AFD, the area of NACK on the constellation should be larger than it of ACK to reduce NACK error rate, which is illustrated in Fig. \ref{Asym sch1}. $\alpha \in \mathbb{R}$ is an index defined by the ratio of the minimum Euclidean distance from the asymmetric decoding threshold to the origin on the constellation diagram over the amplitude. The amplitude of the signal is normalized in Fig. \ref{Const}.
This AFD scheme is distance-based. Other schemes based on other parameters (such as phase shift) are also studied, but they result in poorer performances. We assume that the feedback is transmitted with one binary symbol through an additive Gaussian noise channel (AWGN). For the $i$-th transmission, the ACK error rate $P_{A,i}$ and the NACK error rate $P_{N,i}$ can be computed by:

\begin{equation}
    P_{N,i}=\frac{1}{2}erfc\left((1+\alpha_i)\sqrt{\text{SNR}_f}\right),
\end{equation}
\begin{equation}
    P_{A,i}=\frac{1}{2}erfc\left((1-\alpha_i)\sqrt{\text{SNR}_f}\right),
\end{equation}
where SNR$_f$ is the signal-to-noise ratio of the feedback channel and $\alpha_i$ is the asymmetric detection index of the $i$-th feedback.

\section{Problem formulation}

At the receiver, a message can be considered decodable if the accumulated mutual information (MI) is greater than $N_b$. The probability that the decoding fails after the $i$-th transmission attempt $P_{i,f}$ can be written as:
\begin{equation}
\begin{aligned}
    P_{i,f}&=\mathbb{P}\left\{\sum_{m=1}^{i}I_m< N_b\right\},\\
\end{aligned}
\end{equation}
where $I_m$ is the normalized MI in the $m$-th transmission defined as:
\begin{equation}
    I_m = n_m T_{slot} W\log (1+|h_m|^2\text{SNR}_d),
\end{equation}
where $W$ is the available bandwidth.

To calculate $P_{i,f}$, we need to find the probability density function (pdf) of the accumulated MI. Let $Z_i=\sum_{m=1}^{i} I_l$ denote the summation of MI in the first $i$ transmissions. $P_{i,f}$ can be calculated as $P_{i,f}=\int_{0}^{1}f(z_i)dz_i$, where $f(z_i)$ is the pdf of $Z_i$ computed through $I_1, \dots, I_i$: $f(z_i) = f(I_1)*\dots *f(I_i)$.
To get rid of convolution, the accumulated MI can be approximated by a Gaussian variable which is accurate over low and moderate SNR regimes\cite{Wu}. The approximated pdf of accumulated MI $Z_i$ can be written as:

\begin{equation}
    f(Z_i) = \frac{1}{\sqrt{2\pi\sum_{l=1}^{i}n_lT_{slot}W\sigma_I^2}}e^{-\frac{(Z_k-\sum_{l=1}^{i}n_lT_{slot}W)^2}{2\sum_{l=1}^{i}n_lT_{slot}W\sigma_I^2}},
    \label{pdf}
\end{equation}
where $\bar{I}$ is the mean value of MI and $\sigma_I^2$ is the variance of MI given by \cite{McKay}:

\begin{equation}
    \bar{I}=\log_2(e)e^{\frac{1}{\text{SNR}_d}}\int_{1}^{\infty}t^{-1}e^{-\frac{t}{\text{SNR}_d}}dt
\end{equation}

\begin{equation}
    \begin{aligned}
    \sigma_I^2=\frac{2}{\text{SNR}_d}\log_2^2(e)e^{\frac{1}{\text{SNR}_d}} G^{4,0}_{3,4}\left(1/\text{SNR}_d|_{0,-1,-1,-1}^{0,0,0}\right)-\bar{I}^2,
    \end{aligned}
\end{equation}
where $G^{m,n}_{p,q}\left(^{a_1,\dots,a_p}_{b_1,\dots,b_q}|z\right)$ is the Meijer G-function.

Define by $P_i$ the probability that the $i$-th transmission occurs in a HARQ round, $P_i$ can be calculated via (\ref{Pi}), where we define $P_{N,0}=0$. Accordingly, the overall outage probability $P_{out}$ can be calculated as:

\begin{figure*}[!t]
\normalsize
\begin{equation}
P_i=\left\{
\begin{aligned}
    &1,\:\:\:\:\:\:\:\:\text{if i=1}\\
    &\left(P_{i-1,f}\prod_{j=1}^{i-1}(1-P_{N,j})+\sum_{j=1}^{i-1}\left((P_{i-j-1,f}-P_{i-j,f})\prod_{m=0}^{i-j-1}(1-P_{N,m})\prod_{m=i-j}^{i-1}P_{A,m}\right)\right),\:\:\:\:\:\text{if i=2\dots M}.\\
\end{aligned}
\right.
\label{Pi}
\end{equation}
\hrulefill
\vspace*{-4pt}
\vspace{-1em}
\end{figure*}

\begin{equation}
    P_{out} = 1-\left(P_{1,s}+\sum_{i=2}^{M}\left((P_{i-1,f}-P_{i,f})\prod_{j=1}^{i-1}(1-P_{N,j})\right) \right).
\end{equation}

The average transmission time (service time) of the effective transmissions can be calculated as:

\begin{equation}
    \mathbb{E}[T_{tran}] = \frac{\sum_{i=1}^{M}T_{i}P_{i-1,f}}{\sum_{i=1}^{M}P_{i-1,f}}.
\end{equation}

The total arrival rate (including the packets and the retransmissions) is $\lambda_{tot}=\lambda_0 \frac{\sum_{i=1}^M P_i}{1-P_{out}} $. Therefore, the average waiting and service time in the queue can be calculated as:

\begin{equation}
\begin{aligned}
    \mathbb{E}[T_{queue}] = & \frac{\lambda_{tot}\mathbb{E}[T_{tran}^2]}{2(1-\lambda_{tot}\mathbb{E}[T_{tran}])}\\ = & \frac{\sum_{i=1}^MT_{i}^2P_i}{2\left(\frac{1-P_{out}}{\lambda_0}-\sum_{i=1}^MT_{i}P_i \right)}.
\end{aligned}
\end{equation}

The average service rate has to be greater than the packet arrival rate to make the queuing system stable. Therefore, the follow requirement should always be satisfied:

\begin{equation}
    \frac{\sum_{i=1}^{M}T_{i}P_{i-1,f}}{\sum_{i=1}^{M}P_{i-1,f}} \times \frac{\lambda_0\sum_{i=1}^M P_i}{1-P_{out}} \leq 1.
\end{equation}

According to \cite{38.213}, the UE has to provide corresponding HARQ feedback within $k$ slots after the transmission, where $k$ is specified by the PDSCH-to-HARQ\_feedback timing indicator field in the downlink control information (DCI) format. If we consider the propagation time to be negligible, the total RTT for each transmission is upper bounded by:

\begin{equation}
    T_{tot}  \leq T_{queue}+T_{tran}+kT_{slot}.
\end{equation}

The mean value of the upper bound can be written as:

\begin{equation}
\begin{aligned}
    \bar{T}_{ub} =& \frac{\sum_{i=1}^MT_{i}^2P_i}{2\left(\frac{1-P_{out}}{\lambda_0}-\sum_{i=1}^MT_{i}P_i \right)}\\
    &+ \frac{\sum_{i=1}^{M}T_{i}P_{i-1,f}}{\sum_{i=1}^{M}P_{i-1,f}} + kT_{slot}.
\end{aligned}
\end{equation}

The average delay of the HARQ process can be written in terms of the number of RTTs:
\begin{equation}
    \mathbb{E}[D] = \frac{M+1-\sum_{i=1}^{M-1}P_{i,s}}{1-P_{out}} \times \bar{T}_{ub}.
\end{equation}

In our previous work \cite{AFD}, we have shown that the optimal performance of the HARQ process is achieved when the outage probability is considerably higher than expected. Therefore, the minimum delay with tight outage probability limits $\epsilon$ must be achieved when the actual outage probability $P_{out}=\epsilon$. Then, the optimization problem can be formulated as:

\begin{equation}
\begin{aligned}
    &\min_{n_1,\dots,n_M,\alpha_1\dots \alpha_{M-1}}\:\:\frac{M+1-\sum_{i=1}^{M-1}P_{i,s}}{1-\epsilon} \times\\
    &\left( \frac{\sum_{i=1}^MT_{i}^2P_i}{2\left(\frac{1-\epsilon}{\lambda_0}-\sum_{i=1}^MT_{i}P_i \right)} + \frac{\sum_{i=1}^{M}T_{i}P_{i-1,f}}{\sum_{i=1}^{M}P_{i-1,f}} + kT_{slot} \right)\\
    &\:\:\:\:\:\:\:\:\:\:\:\:\:\:\rm{subject\:to:}\:\:P_{out}\leq \epsilon\\
    &\:\:\:\:\:\:\:\:\:\:\:\:\:\:\:\:\:\:\:\:\:\:\:\:\:\:\:\:\:\:\:\:\:\:\:\:\:\:\frac{\sum_{i=1}^{M}T_{i}P_{i-1,f}}{\sum_{i=1}^{M}P_{i-1,f}} \times \frac{\lambda_0\sum_{i=1}^M P_i}{1-\epsilon} \leq 1\\
    &\:\:\:\:\:\:\:\:\:\:\:\:\:\:\:\:\:\:\:\:\:\:\:\:\:\:\:\:\:\:\:\:\:\:\:\:\:\: n_i \in \mathbb{N}^{+} \cup n_i \leq n_{max},\:\:i=1,2,\dots,M \\
    &\:\:\:\:\:\:\:\:\:\:\:\:\:\:\:\:\:\:\:\:\:\:\:\:\:\:\:\:\:\:\:\:\:\:\:\:\:\:\alpha_k \in \mathbb{R},\:\:k=1,2,\dots,M-1.\\
\end{aligned}
\label{opt2}
\end{equation}

To solve this problem, we need to consider these two sets of variables $\bm{n}=\{n_1, \dots, n_M\}$ and $\bm{\alpha}=\{\alpha_1, \dots, \alpha_{M-1}\}$ separately. In reality, $M$ and $n_{max}$ are not very high values, making the size of the searching space of $\bm{n}$ reasonable. Once $\bm{n}$ is fixed, we can use projected gradient descent (PGD) to find the optimal $\bm{\alpha}$.

\section{Numerical results}

In this section, we compare the long-term average day of our proposed AFD scheme and conventional schemes. Some of the simulation parameters are listed in Table \ref{Simu}. 

\begin{table}[h]
\centering
\caption{Simulation parameters}
\label{tab:my-table}
\begin{tabular}{l|l}
\hline
Parameter                                          & Value \\ \hline
Packet arrival rate  $\lambda_0$                             & 200 packets/\si{\second}\\
PDSCH-to-HARQ\_feedback timing k                                                  & 1     \\
Information length  $N_b$                              & 2816 bits  \\
Slot duration    $T_{slot}$                                  & 125 \si{\micro \second}  \\
Number of transmissions in each HARQ round $M$ & 4     \\ \hline
\end{tabular}
\label{Simu}
\end{table}

\begin{figure}[!t]
	\centering
	\includegraphics[width=0.85\columnwidth]{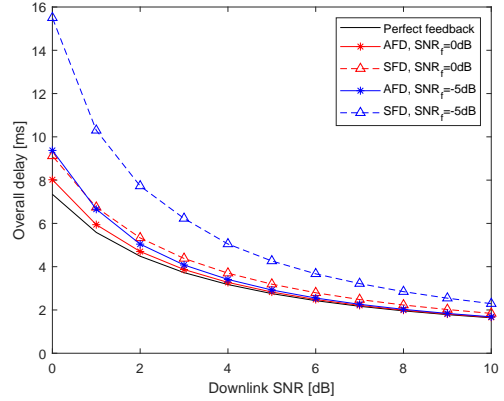}
	\caption{The long-term average delay of our proposed scheme compared with conventional SFD when there is no outage probability limits.}
	\label{unlimited}
\end{figure}

First, we assume that there is no outage probability limits. We compare our proposed AFD scheme with conventional SFD scheme in terms of long-term average delay. Both schemes apply rate adaptation, and even then we can see that our AFD scheme significantly outperforms SFD, especially when the quality of the feedback channel is low. When the downlink channel is of low quality, the average delay of SFD scheme increases sharply, while the impact of low-quality downlink channel on AFD is not significant compared to the system with perfect feedback channel. Compared with a system with perfect HARQ feedback, by applying AFD, we can achieve almost the same performance at SNR$_f=0$dB, and only slightly higher delay at SNR$_f=-5$dB.

\begin{figure}[!t]
	\centering
	\includegraphics[width=0.85\columnwidth]{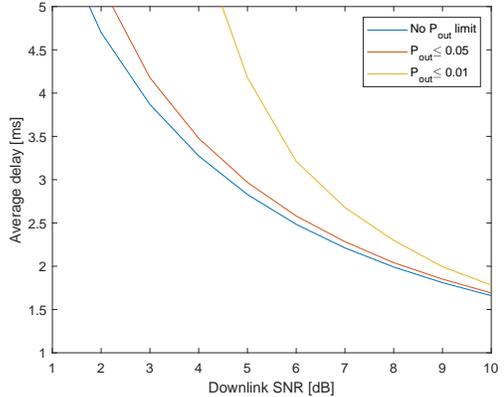}
	\caption{The minimum achievable long-term average delay with outage probability limits when the SNR$_f=0$dB.}
	\label{limited}
\end{figure}

We also find the minimum achievable long-term average delay under strict outage probability limits when SNR$_f=0$dB (see Fig. \ref{limited}). When 5\% outage probability is required, the performance suffers slightly, but still better than the SFD scheme even without the limits. However, when the outage probability is required to be under 1\%, despite it is still achievable with our AFD scheme, the delay is much higher than expected.

\section{Conclusion}

In this work, we optimize the long-term average delay of the HARQ process with imperfect feedback. We analyze different delay components and jointly optimize them by applying AFD and rate-adaptation. The results show that with our proposed scheme, the overall delay of the HARQ process can be significantly reduced, and the impact of low-quality feedback channel can be mitigated. In addition, we can achieve much lower outage probability with AFD at a cost of overall delay.


\vspace{-0.5em}
\bibliographystyle{IEEEtran}
\bibliography{IEEEabrv,MMM}

\end{document}